\documentclass[aps,pre,twocolumn,array,epsfig,eqsecnum]{revtex4}
\usepackage{amsmath}
\usepackage{amsfonts}
\usepackage[dvips]{graphicx}

\begin{document}   
\setlength{\parindent}{0pt}

\title{Restriction on the rank of marginals of bipartite pure states}

\author{S.V.M. Satyanarayana}
\address{
Department of Physics, Pondicherry University, Puducherry 605 014, India}
\date{\today}

\begin{abstract}
Consider a qubit-qutrit ($2 \times 3$) composite state space. Let $C(\frac{1}{2}I_2, \frac{1}{3}I_3)$ be a convex set of all possible states of composite system whose marginals are given by $\frac{1}{2}I_2$ and $\frac{1}{3}I_3$ in two and three dimensional spaces respectively. We prove that there exists no pure state in $C(\frac{1}{2}I_2, \frac{1}{3}I_3)$. Further we generalize this result to an arbitrary $m \times n$ bipartite systems. We prove that for $m < n$, no pure state exists in the convex set $C(\rho_A,\rho_B)$, for an arbitrary $\rho_A$ and rank of $\rho_B >m$.
\end{abstract}

\maketitle

\section{$2 \times 3$ system}
For a two qubit system, Parthasarathy proved that the extremal points of $C(\frac{1}{2}I_2, \frac{1}{2}I_2)$ are maximally entangled pure states \cite{krp}. It was suggested by G.L.Price and S. Sakai \cite{Price} that this is probably true for every $C(\frac{1}{n}I_n, \frac{1}{m}I_m)$. Oliver Rudolf \cite{Oliver} and H Ohno \cite{Ohno} constructed examples of extreme elements of $C(\frac{1}{3}I_3, \frac{1}{3}I_3)$ and $C(\frac{1}{4}I_4, \frac{1}{4}I_4)$ that are not pure states. Recently, Kanmani \cite{Kanmani} has constructed a specific example of an extremal state of $C(\frac{1}{2}I_2, \frac{1}{3}I_3)$ that is mixed entangled state. In this work, we show that there exist no pure states in the convex set $C(\frac{1}{2}I_2, \frac{1}{3}I_3)$.

A general pure state of a qubit qutrit system is given by
\begin{equation}
\rho_{AB}=\sum_{i=0}^1\sum_{j=0}^1\sum_{k=0}^2\sum_{l=0}^2 C_{ik}C^{\star}_{jl}\vert i_Ak_B\rangle \langle j_Al_B \vert
\end{equation}
Marginals are obtained by partial trace over one subsystem. Partial trace over qutrit gives marginal in qubit space, given by
\begin{equation}
\rho_A = \sum_{m=0}^2 \langle m_B \vert \rho_{AB} \vert m_B \rangle
\end{equation}
The partial trace of the general pure state can be shown as
\begin{equation}
\rho_{A}=\sum_{i=0}^1\sum_{j=0}^1\sum_{m=0}^2 C_{im}C^{\star}_{jm}\vert i_A\rangle \langle j_A \vert
\end{equation}
Two choices of coefficients (a) $\vert C_{00} \vert^2 = \vert C_{11} \vert^2 = \frac{1}{2}$ and $C_{01}=C_{10}=C_{02}=C_{12}=0$ and (b) $\vert C_{01} \vert^2 = \vert C_{10} \vert^2 = \frac{1}{2}$ and $C_{00}=C_{11}=C_{02}=C_{12}=0$ result in $\rho_A = \frac{1}{2}I_2$.
On the other hand the partial trace over qubit gives marginal in qutrit space which can be obtained as follows.
\begin{equation}
\rho_B = \sum_{m=0}^1 \langle m_A \vert \rho_{AB} \vert m_A \rangle
\end{equation}
\begin{equation}
\rho_{B}=\sum_{k=0}^2\sum_{l=0}^2\sum_{m=0}^1 C_{mk}C^{\star}_{ml}\vert k_B\rangle \langle l_B \vert
\end{equation}
This marginal can be expanded as follows.
\begin{eqnarray}
\nonumber
\rho_B &=& \left(C_{00}C^{\star}_{00} + C_{10}C^{\star}_{10}\right) \vert 0_B \rangle \langle 0_B \vert  \\
\nonumber
&+& \left(C_{00}C^{\star}_{01} + C_{10}C^{\star}_{11}\right) \vert 0_B \rangle \langle 1_B \vert \\
\nonumber
&+& \left(C_{00}C^{\star}_{02} + C_{10}C^{\star}_{12}\right) \vert 0_B \rangle \langle 2_B \vert \\
\nonumber
&+& \left(C_{01}C^{\star}_{00} + C_{11}C^{\star}_{10}\right) \vert 1_B \rangle \langle 0_B \vert \\
&+& \left(C_{01}C^{\star}_{01} + C_{11}C^{\star}_{11}\right) \vert 1_B \rangle \langle 1_B \vert \\
\nonumber
&+& \left(C_{01}C^{\star}_{02} + C_{11}C^{\star}_{12}\right) \vert 1_B \rangle \langle 2_B \vert \\
\nonumber
&+& \left(C_{02}C^{\star}_{00} + C_{12}C^{\star}_{10}\right) \vert 2_B \rangle \langle 0_B \vert \\
\nonumber
&+& \left(C_{02}C^{\star}_{01} + C_{12}C^{\star}_{11}\right) \vert 2_B \rangle \langle 1_B \vert \\
\nonumber
&+& \left(C_{02}C^{\star}_{02} + C_{12}C^{\star}_{12}\right) \vert 2_B \rangle \langle 2_B \vert
\end{eqnarray}
For $\rho_B = \frac{1}{3}I_3$, we must set the coefficients of $\vert k_B \rangle \langle l_B \vert$ with $k \neq l$ to zero and the coefficient of the same term with $k=l$ has to be set to 1/3. It can be see that such a choice does not exist.

Here is another way to understand the result and the following proof is given by S. Kanmani \cite{Kanmani2}. Let $\rho_{AB}$ be a pure state density matrix of $2 \times 3$.
Since, a pure state is a rank-one projection operator, $\rho_{AB}$ can be written as $ V^{*} V $, where $V= (a_1, a_2, ..., a_6)$
is a  row vector and $V^{*}$  is the conjugate transpose of it. Hence,\\

\begin{equation}
\rho_{AB}=\left(
  \begin{array}{cccccc}
 \bar{a_1}a_1 & \bar{a_1}a_2 &  \bar{a_1}a_3 & \bar{a_1} a_4 & \bar{a_1}a_5 & \bar{a_1}a_6\\
\bar{a_2}a_1 & \bar{a_2}a_2 &  \bar{a_2}a_3 & \bar{a_2} a_4 & \bar{a_2}a_5 & \bar{a_2}a_6 \\
\bar{a_3}a_1 & \bar{a_3}a_2 &  \bar{a_3}a_3 & \bar{a_3} a_4 & \bar{a_3}a_5 & \bar{a_3}a_6 \\
\bar{a_4}a_1 & \bar{a_4}a_2 &  \bar{a_4}a_3 & \bar{a_4} a_4 & \bar{a_4}a_5 & \bar{a_4}a_6 \\
\bar{a_5}a_1 & \bar{a_5}a_2 &  \bar{a_5}a_3 & \bar{a_5} a_4 & \bar{a_5}a_5 & \bar{a_5}a_6 \\
\bar{a_6}a_1 & \bar{a_6}a_2 &  \bar{a_6}a_3 & \bar{a_6} a_4 & \bar{a_6}a_5 & \bar{a_6}a_6
\end{array}
\right)
\end{equation}
This allows one to re-write the $\rho_{AB} $ in the form of a block matrix of $ 2\times 2$, where each
entry is a $3 \times 3 $ scalar matrix. \

\begin{equation}
\rho_{AB}=\left[
  \begin{array}{cc}
  V_{1}^{*} V_1 &  V_{1}^{*} V_2 \\

V_{2}^{*} V_1 &  V_{2}^{*} V_2
  \end{array}
\right]
\end{equation}
Where $ V_1 = (a_1, a_2 ,a_3) $ and $ V_2 = (a_4,a_5,a_6) $ and hence $V_{1}V^{*}_{1} $ and $V_{2}V^{*}_{2}$ are both $ 2 \times 2$ scalar matrices of rank one.  It can be verified that, $\rho_{B}$, one of the marginal states
of $\rho_{AB} $ is given as $\rho_B = V_{1}^{*} V_1 + V_{2}^{*} V_2 $. Thus, $\rho_B$ is sum of two rank-one matrices.
Since, rank( A + B) $\leq $ rank(A) + rank(B), it follows that the rank of $\rho_B$ has to be less than or equal to
two. As $\frac{1}{3}1_{3}$ is a matrix of rank three, it follows that there exist no pure state in $ 2 \times 3 $ whose marginal is
$\frac{1}{3}1_{3}$. The following proposition is a generalization of this observation.\\

{\bf Proposition } Consider   a composite quantum system $H_A\otimes H_B$, of dimension $ m \times n $.
Then, the rank of the marginal $\rho_{B}$, that corresponds to that any pure state $\rho_{AB} $ of $H_A\otimes H_B$, has to be less than or equal to m.\\

This is because, $\rho_B$ is sum of m number of rank one matrices and hence cannot have a rank more than m. Clearly,
when $m \geq n $ this is a trivial assertion. However, when $ m < n $ the above proposition provides some useful information as in
the context of the convex set of $C(\rho_A, \rho_B)$.\\

{\bf Proposition } Consider the convex set of $ C(.\ , \rho_B)= \{ \rho_{AB} : Tr_{A}(\rho_{AB})= \rho_{B} \} $ in the context of
a composite quantum system $H_A \otimes H_B $ of dimension $m \times n $ . If m is less than the rank of $\rho_B$ then the set
$ C(.\ , \rho_B)$  does not contain any pure state.

Parthasarathy \cite{krp} proved that there is a upper bound for the the rank of the extreme
 elements of the convex set $ C(\rho_A, \rho_B) $. Similarly, when the composite system is $ C^m \otimes C^n $ , where $m < n$,  the rank of the extreme elements of $ C(\rho_A, \rho_B) $ has to be more than one if rank of $\rho_B > m$.


\begin{thebibliography}{100}
\bibitem{krp}
K.R.Parthasarathy, Extremal quantum states in coupled quantum systems with fixed marginals, {\it Ann. Inst. H. Poincare}, \textbf{41} 257-268 (2005).
\bibitem{Price}
G.L.Price and S.Sakai, Extremal marginal tracial states in coupled systems, {\it Operators and matrices}, \textbf{1} 153-163 (2007).
\bibitem{Oliver}
O.Rudolph, On extremal quantum states of composite systems with fixed marginals, {\it J. Math. Phys.}, \textbf{45} 4035-4041 (2004).
\bibitem{Ohno}
H. Ohno, Maximal rank of extremal marginal tracial states, {\it J. Math. Phy.}, \textbf{51} 092101-092110 (2010).
\bibitem{Kanmani}
S. Kanmani, A note on extremal states of composite quantum systems with fixed marginals, arXiv:1304.5413v1 (quant-ph) (2013).
\bibitem{Kanmani2}
S. Kanmani, {\it Private Communication}.
\end{thebibliography}
\end{document}